\documentclass[10pt]{article}

\usepackage{amsmath}
\usepackage{amssymb}

\usepackage{graphicx}

\usepackage{cite}

\usepackage{color,nicefrac,subfigure,booktabs} 

\usepackage[labelfont=bf,labelsep=period,justification=raggedright]{caption}

\makeatletter
\renewcommand{\@biblabel}[1]{\quad#1.}
\makeatother

\date{}

\begin{document}

\begin{flushleft}
{\Large
\bf{iSeg: an algorithm for segmentation of genomic data}\linebreak
}
\vspace{1cm}
\linebreak
S.B. Girimurugan$^{1\P \ast}$, Jonathan Dennis$^{2\&}$, Jinfeng Zhang$^{3\P\ast}$\linebreak
\vspace{0.1cm}

{1} Department of Mathematics, Florida Gulf Coast University, Fort Myers, FL, USA
\vspace{0.1cm}

{2} Department of Biological Science, Florida State University, Tallahassee, FL, USA
\vspace{0.1cm}

{3} Depart1ent of Statistics, Florida State University, Tallahassee, FL, USA
\vspace{0.5cm}

$\ast$ E-mail:sgirimurugan@fgcu.edu\\
\vspace{0.1cm}

$\P$ These authors contributed equally to this work. \\
\vspace{0.1cm}
$\&$ This author also contributed equally to this work.
\vspace{0.1cm}
\end{flushleft}

\section*{\huge Abstract}
Identification of functional elements of a genome often requires dividing a sequence of measurements along a genome into segments differing from adjacent segments. In many applications, the mean of the measured values at multiple genomic locations in a segment is used to make
inference of the property of interest. The segments with non-zero means often correspond to genomic regions with certain biological events, such as changes between two conditions. This problem is often
called the segmentation problem in the field of genomics, and the change-point problem in other scientific disciplines. We designed an efficient algorithm, called iSeg, for segmentation of high-throughput
genomic profiles. iSeg first utilizes dynamic programming to compute the significance for a large number of candidate segments. It then uses tree-based data structures to detect overlapping significant regions and update them simultaneously. Refinement and merging of significant segments are performed at the end to generate the final segmentation. We evaluate iSeg using both simulated and experimental datasets and show that it performs quite well when compared with existing methods.
\section*{\huge Introduction}
High throughput experiments, such as microarray
and sequencing, are powerful tools for studying genetic and epigenetic
functional elements at genome scale \cite{encode2012integrated}.
There has been a large number of studies on the
analysis of gene expression data generated from high-throughput experiments
\cite{KhatriSB122}. When measuring gene expressions,
the genomic locations of genes are known and multiple probes (or short
reads) can be mapped to a gene to obtain its expression values. With
replicates from two experimental conditions, standard hypothesis tests,
such as t-test, can be performed to infer the differentially expressed
genes. On the other hand, the situation is quite different for functional
elements without predefined locations (i.e. starting and ending positions).
Consider DNA copy number as an example. When detecting the changes in
DNA copy number between two experimental conditions, one needs to
consider a very large number of regions that can possibly undergo
changes. The number is usually much larger than the total number of
genes. Other functional elements especially epigenetic features fall
into the same category. This poses a significant challenge to the
analysis of such type of data. 

The problem is usually formulated as segmenting
a sequence of measurements along the genome. For example, if segments
without changes have a mean value of zero and those with changes have
nonzero means, then the goal is to identify those segments of the
genome whose means are significantly above or below zero. 
A number of methods have been developed recently and many of those were tested
on analysis of DNA copy number variations (CNVs) for microarray-based comparative genomic hybridization (aCGH) data \cite{pmid15475419,
jstor4541408, rancoita2009bayesian, zhang2007modified, Diskin2006, pmid17234643, baldi1998machine, BaldiL01, HMMSeg, PicardLBR11, jeng2010optimal, tcai, wang2007penncnv}. The previous methods fall into several categories including change-point detection \cite{pmid15475419, pmid17234643, sen1975tests, baldi1998machine, BaldiL01, rancoita2009bayesian, Picard2005, PicardLBR11, chen2009statistical, chen2011bayesian, Niu2012, Yao1993,killick2011changepoint, cleynen2013segmentor3isback}, Hidden Markov models \cite{HMMSeg,wang2007penncnv,Marioni2006, Stjernqvist2007,jaschek2009spatial}, Dynamic Bayesian Network (DBN) models \cite{hoffman2012unsupervised,hoffman2012integrative}, signal smoothing \cite{hupe2004analysis, ben2008fast, Tibshirani2008, hu2007exploiting}, and variational models \cite{nilsson2009ultrasome, Morganella2010}. For review and comprehensive comparison, please refer to \cite{kharchenko2008design, park2008experimental, lai2005comparative, willenbrock2005comparison}. In recent years, many efforts have been focused on developing methods for segmentation of multiple profiles simultaneously \cite{zhang2012reconstructing, zhou2013multisample, Zhang2010, baladandayuthapani2010bayesian,Pique-Regi2009, Wiel2009, Picard2011a, Diskin2006, Guttman2007, beroukhim2007assessing, Shah2007,Zhang2010b,Zhang2010a,Nowak2011}. 
Despite significant progresses made in this area, further improvement in terms of both accuracy and computational speed is still desirable. In addition, some methods require
users to adjust parameters to obtain acceptable results. 
In this study, we designed an algorithm to segment
genome-wide profiles to achieve better accuracy and efficiency
compared to existing methods. In addition, we minimize the number
of parameters users have to tune so that our method can be easily
applied by biologists with limited analytical expertise. Our method,
iSeg (implemented in C++), has shown superior performance on both simulated data and benchmark
experimental data compared with previous methods. The next section
describes the method in detail. 
\section*{\huge Materials and Methods}
\begin{figure}
\begin{center}
\includegraphics[width=0.95\columnwidth]{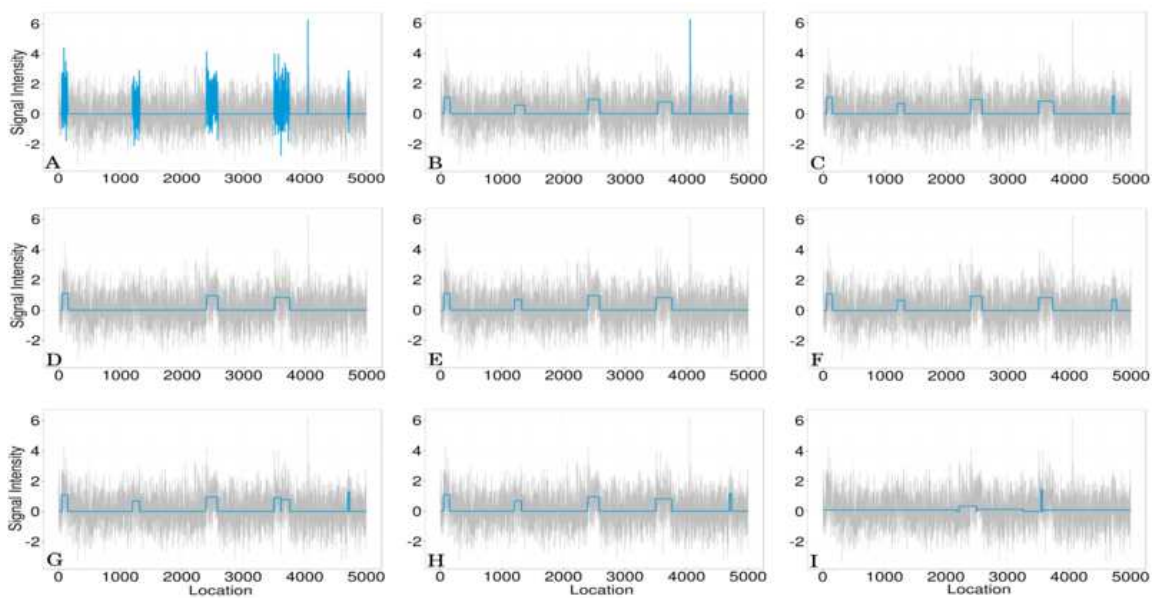}
\end{center}
\caption{\textbf{One of the simulated profiles and its detected segments obtained using iSeg.} (A) The actual data with background noise and meaningful segments. The segments with non-zero means are normally distributed with unit variance and means 0.72, 0.83, 0.76, 0.9, 0.7, and 0.6 respectively. The profiles shown here are normalized for an approximate signal to noise ratio of 1.0. The segments detected by iSeg (B) and other existing methods: snapCGH (C), mBPCR (D), cghseg (E), cghFLasso (F), HMMSeg (G), DNAcopy (H) and fastseg (2) Comparison of $F_{1}$-scores for the simulation profiles with SNR$\simeq$1.0. \color{black} Since the profiles are simulated, the SNR of the resultant profiles is approximately one. The SNR defined during the simulation is exactly one. \color{black} } 
\label{fig:Sim}
\end{figure}

Most segmentation methods have an underlying assumption
of normality. For instance, the test statistics in \cite{pmid15475419, BaldiL01, jeng2010optimal} are modified versions
of a t-statistic. We make a similar assumption in this study, so the comparison with existing methods is straightforward. 
\\
Consider a sample consisting of $N$ measurements along the genome in a sequential order, $\textrm{X}_{1},\textrm{X}_{2},\ldots,\textrm{X}_{N}$, and
\begin{eqnarray}
X_k &\sim & N(0,\sigma^2), \quad\quad\quad \forall k \in \mathcal{G}\\
X_k &\sim & N(\mu_i,\sigma^2), \quad\quad\,\,\,\, \forall k \not\in \mathcal{G}
\end{eqnarray}
for some set of locations `$\mathcal{G}$'.
The common assumption is that there are $M$ non-overlapping segments
with mean $\mu_{1},\mu_{2},\ldots,\mu_{i},\ldots,\mu_{M}$, where
$\mu_{i}\neq0$, and the union of these segments will form the complement of the set `$\mathcal{G}$'. \color{black}If the background level is non-zero, tests can be performed over segments for this level instead of testing against zero. According to this model, it is possible for multiple segments with different non-zero means to be adjacent to each other. In addition, all the measurements are assumed to be independent. \color{black}       This assumption has been employed in many existing methods \cite{pmid15475419,Picard2005}. A summary of existing methods that use such an i.i.d assumption and its properties are nicely discussed in \cite{roy2013evaluation}.  \color{black} 
The goal of a segmentation method is to detect all the $M$ segments with non-zero means. \\
A formal description of change-point problems is given in \cite{brodsky1993nonparametric}. \color{black} As an example, (Fig. \ref{fig:Sim}(A)) shows segments sampled from Normal distributions with non-zero means where the rest of the data is sampled from a standard Normal distribution. \color{black} One approach used by some of the previous methods \cite{jeng2010optimal, pmid15475419} is to first find a segment with the highest significance (or smallest  p-value), remove the segment and repeat the process for the rest of the profile until all the segments with significance higher than a threshold value are identified.
There are two computational challenges associated with this approach that also manifest in many previous methods. First, the number of segments that have to be examined is very large; Second, the overlaps among significant segments need to be detected so that the significance of the overlapping segments can be adjusted accordingly. To deal with the first challenge, we applied dynamic programming combined with exponentially increased segment scales to speed up the scanning of a large sequence of data points. \color{black} The resulting optimization approach is top-down with memoized computations identifying optimal substructures. \color{black} To deal with the second challenge, we designed an algorithm coupling two balanced binary trees to quickly detect overlaps and update the list of most significant segments. Segment refinement and merging allow iSeg to detect segments of arbitrary length. The details of the algorithm are given below.
\\
\subsection*{\large Computing p-values using dynamic programming} iSeg scans a large number of segments starting with a minimum window length, $W_{min}$, and up to a maximum window length, $W_{max}$. They have default values 1 and 300, respectively. \color{black} This window length increases by a fixed multiplicative factor, called power factor ($\rho$), with every iteration. \color{black} For example, the shortest window length is $W_{min}$, and the next shortest window length would be $\rho W_{min}$. The default value
for $\rho$ is 1.1. When scanning with a particular window length, \textit{W}, we use overlapping
windows with a space of \textit{W/5}. \color{black} When `W' isn't a multiple of 5, numerical rounding (\textbf{ceil}) is applied. The aforementioned parameters can be changed by a user. \color{black} The algorithm computes p-values for candidate segments and detects a set of non-overlapping segments most significant among all possible segments. 
\\
Given the normality assumption, a standard test for mean is the one-sample
student's t-test, which is commonly found among many existing methods. The test statistic for this test is,
\begin{eqnarray}
t &=& \dfrac{\bar{x}\sqrt{n}}{s}
\end{eqnarray}
where $\bar{x}$ is the sample mean, $s$ is the sample standard deviation, and $n$ is the sample size. A drawback of this statistic is that it cannot evaluate segments of length 1. This may be the reason that some of the previous methods are not good at detecting segments of length 1. Although we can derive a test statistic separately for segments of length 1, the two statistics may not be consistent. To solve this issue, we first estimate the sample standard deviation using median absolute deviation and assume that the standard deviation is known. This allows us to use $z$ statistic instead of $t$ statistic and the significance of single points can be evaluated based on the same model assumption as longer segments.
To calculate sample means for all segments to be considered for significance, the number of operations required by a brute force approach is `$C_b$'.
\begin{eqnarray}
C_b &=& \sum_{i=0}^{k} (N-\rho^i W_{min})\rho^{i} W_{min}
\end{eqnarray}
where, $\rho^{k}W_{min}\leq W_{max}$ and $\rho^{k+1}W_{min}>W_{max}$.\\
Computation of these parameters (means and standard deviations) for larger segments can be made more efficient by using the means computed for shorter segments \color{black}(top-down memoization of dynamic programming). For example,the running sum of a shorter segment of length `m' given by,
\begin{eqnarray}
S_{m}= \sum_{i=1}^{m}X_{i},
\end{eqnarray}
If this sum is retained, the running sum of a longer segment of length `r'($>m$) in the next iteration can be obtained as,
\begin{eqnarray}
S_r=S_m+\sum_{i=(m+1)}^{r} X_i
\end{eqnarray}
and the means for all the segments can be computed using these running sums. Now, the total number of operations ($C_b^{*}$) is
\begin{eqnarray}
C_b^{*} &=& N+\sum_{i=0}^{k}(N-\rho^i W_{min})
\end{eqnarray}
which is much smaller in practice than the number of operations ($C_b$) without
using dynamic programming. Computation of standard deviations are sped up using a similar memoization process resulting in computational efficiency.\color{black}
 \\
\subsection*{\large Detecting overlapping segments and updating
significant segments using balanced binary trees}
When the p-values of all the segments are computed, we rank the segments
by their p-values from smallest to largest. All the segments with
p-values smaller than a threshold value, $p_{s}$, are kept in a balanced
binary tree (BBT1). We set $p_{s}$ as 0.001 in this study to speed
up the computation. With thousands of hypothesis tests being performed
usually for a particular dataset, this cutoff is reasonable. \color{black} Assuming  a significance level ($\alpha$) of 0.1, 100 simultaneous tests will maintain a family-wise error rate (FWER) bounded by 0.001 with Bonferroni and Sidak corrections. Thus, the cut-off is an acceptable upper bound for multiple testing. \color{black}It can
be changed by a user if necessary. The procedure is described below
as a pseudo-code. The set SS stores all significant segments. The
second balanced binary tree (BBT2) stores the boundaries for significant
segments. After the procedure, SS contains all the detected significant segments. The selection of segments using balanced binary tree aims to minimize the p-values for individual segments. When the minimization causes some segments to overlap, the one with smaller p-value is selected.
\\
\begin{center}
\framebox{
\begin{minipage}[t]{0.925\columnwidth}
\textbf{procedure SelectSignificantSegments}\\
initialize BBT2 // BBT2 is empty at the beginning\\
while(BBT1 not empty)

\qquad$S$ = top ranked segment in BBT1 (smallest
p-value among all segments in BBT1)\\

{\qquad{}delete S from BBT1} \\

{\qquad{}$l$ = left boundary of $S$} \\

{\qquad{}$r$= right boundary of $S$} \\

{\qquad{}if(checkoverlap (BBT2, $l$, $r$) == FALSE)
// no overlapping} \\

{\qquad{}\qquad{}insert pair($l$, $r$) into BBT2} \\

{\qquad{}\qquad{}insert $S$ to set SS} \\
\end{minipage}}
\end{center}
\subsection*{\large Refinement of significant segments} The significant segments are refined further by expansion and shrinkage. Without loss of generality, in the procedure (see SegmentExpansion text box) we describe expansion on left side of a segment only. Expansion on the right side and shrinkage are done similarly. When performing said expansion and shrinkage,  a condition to check for overlapping segments is applied so the algorithm results in only disjoint segments.
\hfill \\
\subsection*{\large Merging adjacent significant segments}
When all the significant non-overlapping segments are detected and
refined in the previous steps, iSeg performs a final merging step
to merge adjacent segments. The procedure is straightforward. We check
each pair of adjacent segments. If the merged segment, whose range
is defined by the left boundary of the first segment and the right
boundary of the second segments, has a p-value smaller than those
of individual segments, then we merge the two segments. The new segment
will then be tested for merging with its adjacent segments iteratively.
The procedure continues until no segments can be merged. \color{black} Apart from refinement, merging also prevents partitioning of a signal into only short segments. Short segments are called significant only if a longer segment or several merged segments are insignificant. With refinement and merging, iSeg can detect segments of arbitrary length--- long and short. \color{black}
\\
\begin{center}
\framebox{
\begin{minipage}[h]{0.925\columnwidth}%
\textbf{{procedure SegmentExpansion (}}$S_{l,r}$\textbf{{)}}

/{*} $S_{l,r}$: the segment to be expanded. Its left boundary is
$l$, and right boundary is $r$. {*}/

$S$ = $S_{l,r}$

{while()}

{\qquad{}}\textit{{p}}{{}
= p-value of $S$}

{\qquad{}}\textit{{L}}{{}
= length of $S$}

{\qquad{}}\textit{{$l_{0}$}}{{}
= left boundary of $S$}

{/{*} expand the segment by 1/}\textit{{K}}{{}
of its current length, and compute its p-value. The default value
for }\textit{{K}}{{} is 10. {*}/}

{\qquad{}$l'=l_{0}-ceiling(L/K)$ }

{\qquad{}$p'$ = p-value of segment }$S_{l',r}$

{\qquad{}if $p' < p$}

{\qquad{}\qquad{}$S$ = }$S_{l',r}$

{\qquad{}else}

{\qquad{}\qquad{}compute p-values for all segments
with left boundary in $(l',l_{0})$ and right boundary $r$. }

{\qquad{}\qquad{}let $p_{m}$ be the minimum p-value
of these segments, and $l_{m}$ be the corresponding left boundary}

{\qquad{}\qquad{}if $p_{m}<p$}

{\qquad{}\qquad{}\qquad{}$S$ = }$S_{l_{m},r}$

{\qquad{}\qquad{}break}

Update $S_{l,r}$with boundaries of $S$.%
\end{minipage}}
\end{center}
\subsection*{\large Multiple comparison}
In iSeg, p-values for potentially significant segments are calculated.
Using a common p-value cutoff, for example 0.05, to determine significant
segments can suffer from a large number of false positives due to
multiple comparison. To cope with the multiple comparison issue which
can be very serious when the sequence of measurements is long, we
use a false discovery rate (FDR) control. Specifically, we employ
the Benjamini-Hochberg (B-H) procedure \cite{benjamini1995controlling}
to obtain a cutoff value for a predefined false discovery rate $\left(\alpha\right)$,
which has a default value of 0.01, and can also be set by a user.
Other types of cutoff values can be used to select significant segments,
such as a fixed number of most significant segments.
\subsection*{\large Biological cutoff} Often in practice, biologists prefer to call signals above a certain threshold.
For example, in gene expression analysis, a minimum of two-fold change
may be applied to call differentially expressed genes. Here we add
a parameter, $p_{b}$, which can be tuned by a user to allow more
flexible and accurate calling of significant segments. \color{black} Such a cutoff is quite useful in situations where the baseline is non-zero.\color{black}

\section*{\huge Results}
We compare our method with several previous methods
for which we can obtain executable programs: HMMSeg \cite{pmid17384021},
CGHSeg \cite{PicardLBR11}, DNAcopy \cite{sen1975tests, pmid15475419}, fastseg \cite{BaldiL01}, cghFLasso \cite{Tibshirani2008}, BioHMM-snapCGH \cite{Marioni2006} and mBPCR \cite{rancoita2009bayesian}. Each method has some parameters
that can be tuned by a user to achieve better performance. In our
comparative study, we carefully chose the parameters based on the
recommendations provided by the authors of such
methods. For each method, a single set of parameters is used for all
data sets. Post-processing is required by some of the methods to identify
significant segments. 

In our analysis, performance is measured using $F_1$ -scores \cite{Chinchor92}
for all methods. $F_1$ -scores are considered as a robust measure
for classifiers, as they account for both precision and recall in
their measurement. The $F_1$ -score is defined as,
\begin{eqnarray}
 F_1=\nicefrac{\left(2pr\right)}{\left(p+r\right)}
\end{eqnarray}
where $p$ is precision and $r$ is recall for a classifier. In terms
of the true and false positives, 
\begin{eqnarray}
p &=&\nicefrac{TP}{\left(TP+FP\right)}\\
r &=&\nicefrac{TP}{\left(TP+FN\right)}. 
\end{eqnarray}
Since an effective threshold is not varied to assess performance in the space of sensitivity and specificity, our analysis implements $F_1$-scores. These scores have been shown to effectively assess the performance of various classifiers (including support vector machines) and are claimed as viable alternatives to approaches such as ROC, AUC etc.\cite{sokolova2006beyond}.

The methods CGHSeg, DNAcopy, and fastseg depend on random seeds given
by a user (or at run-time automatically), and the $F_{1}$-scores at
different runs are not the same albeit very close.
These methods were run using three different random seeds, and the
averages of the $F_{1}$-scores were used to measure their performance.
In the following sections, we assess iSeg's performance using both
simulated data and experimental benchmark data. 

\subsection*{\large Performance on simulated data}
The simulation profiles were generated under varying
noise conditions, with signal to noise ratios (SNR) of 0.5, 1.0 and
2.0, which correspond to poor, realistic and ideal cases. Ten different
profiles of length 5000 are simulated. 

\begin{figure}
\begin{centering}
\includegraphics[width=0.95\textwidth , height=8cm]{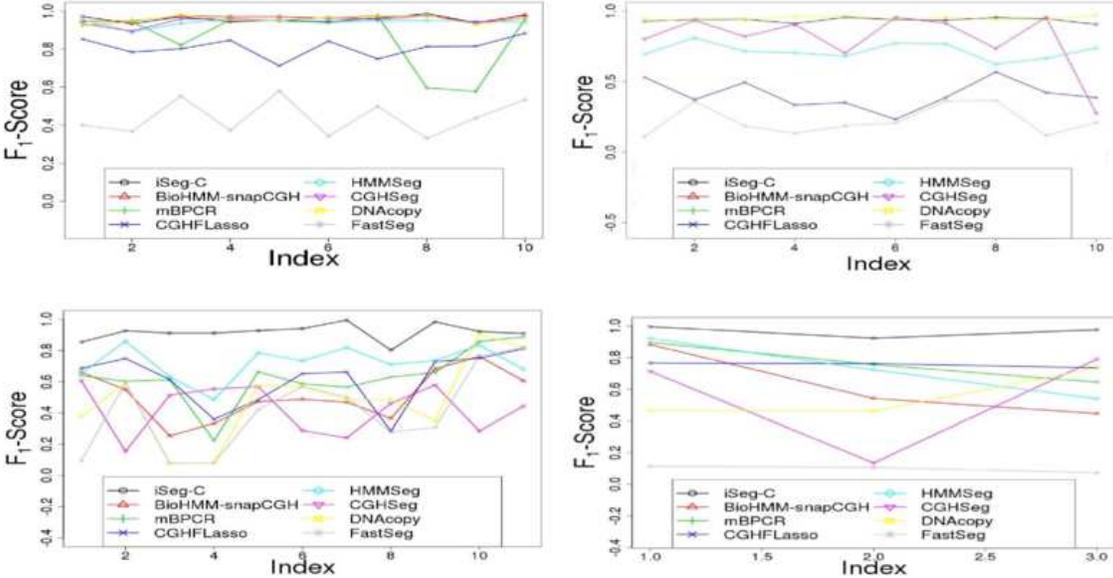}
\caption{\textbf{Comparison of $F_{1}-$scores of various methods in analyzing different types of profiles.} A. Simulated Profiles B. Simulated Long Profiles (n=100K) C. Coriell (Snijders et al.) Profiles D. BACarray Profiles \label{fig:F1scores}}
\end{centering}
\end{figure}

For each profile, five different segments of varying lengths are predefined at different locations.
Data points outside of these segments are generated from normal distribution
with mean zero. The five segments are simulated with non-zero means
and varying amplitudes (some easy to detect and some rather difficult)
in order to assess the robustness of the methods. (Fig. \ref{fig:Sim})
shows an example of the simulated data and the segments identified
by iSeg and other existing methods.
(Fig. \ref{fig:F1scores}(A)) shows the performance of iSeg and other methods on simulated data with
SNR = 1.0. We can see that iSeg, DNACopy and CGHSeg perform similarly well, with HMM and CGHFLasso performing a little worse while fastseg did not perform as well as the other methods.
iSeg is also tested using a set of 10 longer simulated profile, each of length 100000.
Seven segments are introduced at varying locations along the profiles.
iSeg performs still quite well in these very long profiles. The performance of these methods on long sequences is shown in (Fig. \ref{fig:F1scores}(B)).

\subsection*{Performance on experimental data}

To assess the performance of iSeg on experimental
data, we use three different datasets: 11 profiles from \cite{snijders2001assembly},
called Coriell dataset; three profiles from \cite{wang2005method}
, called BACarray dataset; and two profiles taken from TCGA (The Cancer Genome Atlas), called TCGA dataset. The 11 profiles in Coriell datasets correspond to 11 cell lines: GM03563, GM05296, GM01750, GM03134, GM13330, GM01535, GM07081, GM13031, GM01524, S0034 and S1514. We construct ``gold standard" annotations using a consensus approach. We first run all the methods using several different parameter settings for each method. The resulting segments are evaluated using the test statistic described in this method. The set of gold standard segments are obtained using Benjamini-Hochberg procedure to account for multiple comparisons. The annotations derived using a consensus approach are provided as Supplementary material. 

\begin{figure}
\begin{center}
\includegraphics[width=0.95\columnwidth,height=6cm]{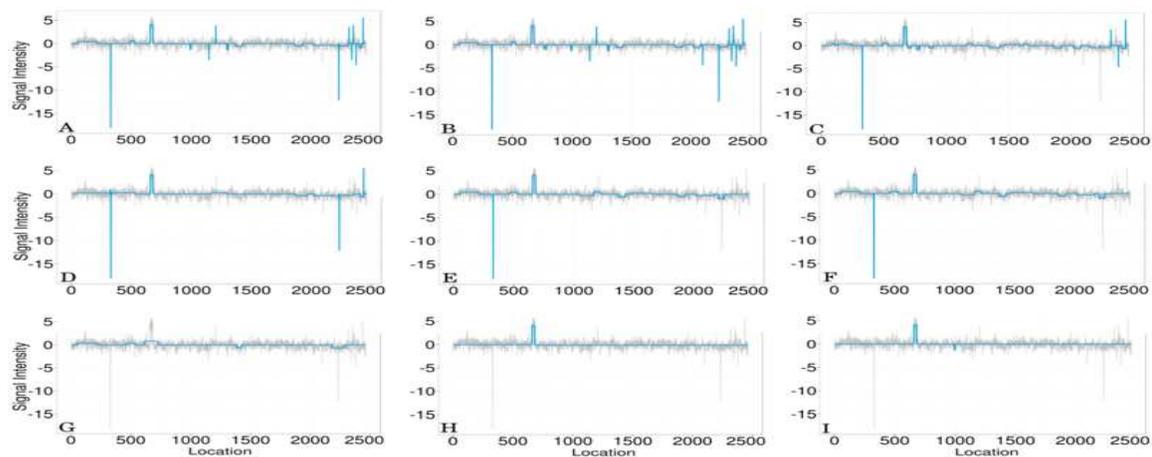}

\caption{\textbf{Comparison of segmentation results obtained for the Coriell dataset.} (A) The gold standard segmentation obtained using a  consensus approach. Segmentation results of iSeg (B) and other existing methods: snapCGH (C), mBPCR (D), cghseg (E), cghFLasso (F), HMMSeg (G), DNAcopy (H) and fastseg (I).\label{fig:coriell_seg}}
\end{center}
\end{figure}

\begin{figure}
\begin{center}
\includegraphics[width=0.95\columnwidth,height=6cm]{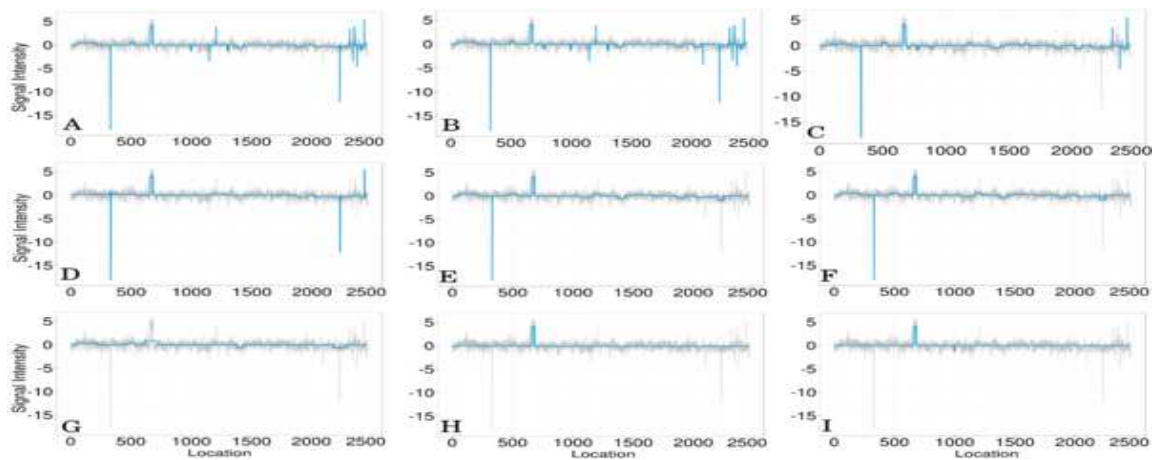}
\caption{\textbf{Comparison of segmentations for BACarray dataset.} (A) The gold standard segmentation obtained using a  consensus approach for the TCGA Glioblastoma Multiforme (GBM) profile. Segmentation results of iSeg (B) and other existing methods: snapCGH (C), mBPCR (D), cghseg (E), cghFLasso (F), HMMSeg (G), DNAcopy (H) and fastseg (I).\label{fig:BAC_seg}}
\end{center}
\end{figure}

The 11 profiles from the Coriell dataset were segmented using iSeg and the other methods, and the $F_{1}$-scores are shown in (Fig. \ref{fig:F1scores}(C))
%
The performance of iSeg is robust with accuracy above 0.75 for all the profiles from this dataset and it is comparable to, better in some cases, other methods. For HMMSeg, both no-smoothing and smoothing are used. The best smoothing scale for HMMSeg was found to be 2 for the Coriell dataset. The segmentation results for one profile in Coriell dataset is shown in (Fig. \ref{fig:coriell_seg}).
We can see that iSeg identified most of the segments. While DNAcopy, fastseg, HMMSeg and cghseg missed single-point peaks, cghFLasso, mBPCR and snapCGH missed some longer segments. 
The segmentation results for other profiles in Coriell dataset can also be found in the Supplementary material.
We generated annotations using the consensus method for BACarray dataset similar to the Coriell dataset. The comparison of F1-scores is shown in (Fig. \ref{fig:F1scores}(D)) and the comparison of segmentation results is shown in (Fig. \ref{fig:BAC_seg}). iSeg has better $F_{1}$-scores than the other methods and a similar conclusion can also be gathered from visual inspection.

\begin{figure}[h!]
\begin{center}
\includegraphics[width=0.95\columnwidth ,height=9cm]{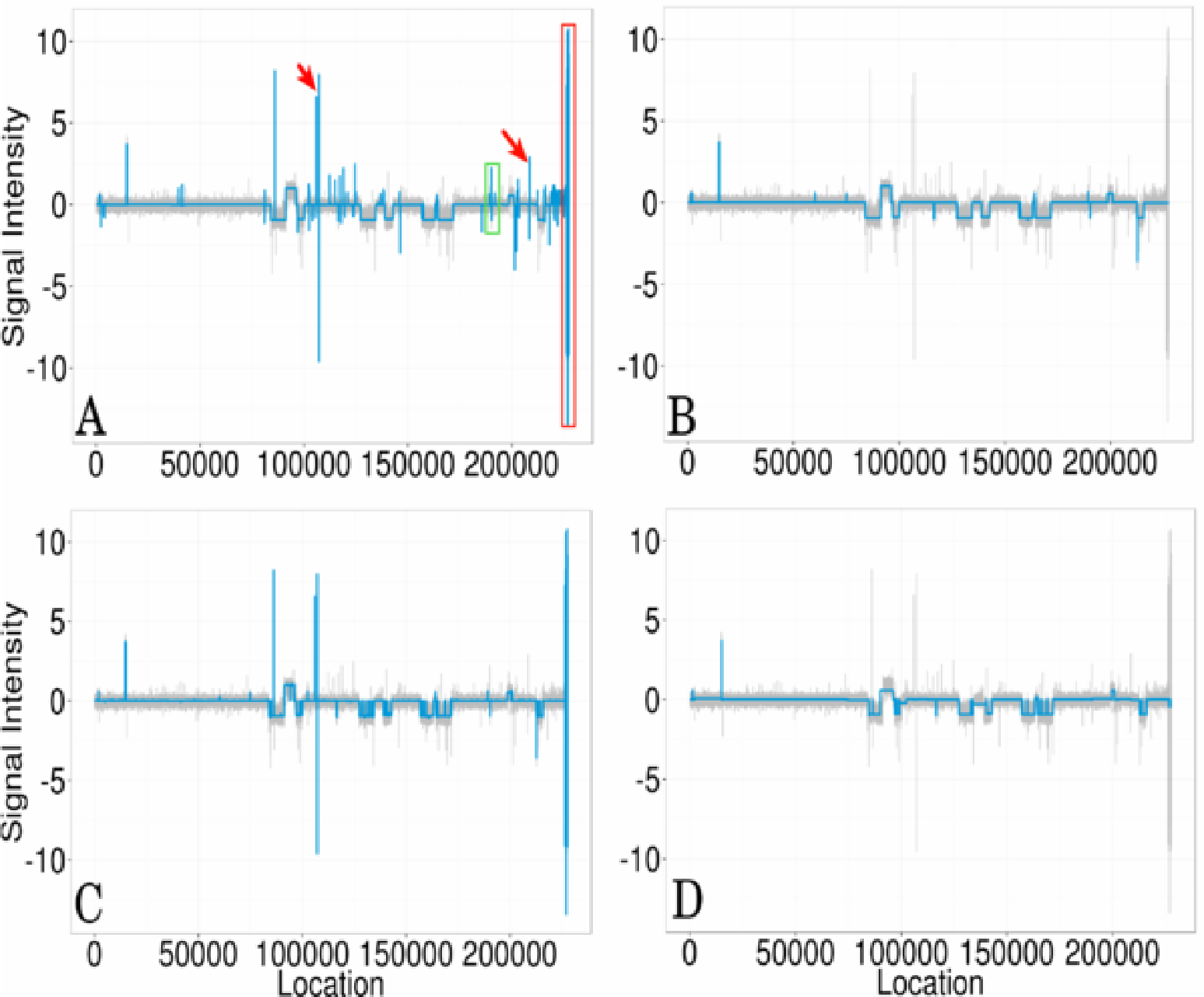}
\end{center}
\caption{ \textbf{Comparison of segmentations for the TCGA dataset.}  The patient profile ID is TCGA-02-0007 and the data is supplied by the Harvard Medical School 244 Array CGH experiment (HMS). Segmentation results of iSeg (A) and other existing methods: DNAcopy (B), cghFLasso (C), and cghseg (D). The peaks pointed by the arrows and the region labeled by the red, and green squares are identified by iSeg, but not all of them are detected by the other three methods. Overall, iSeg consistently identifies all the significant peaks. Other methods often miss peaks or regions which are more significant than those identified.   \label{fig:TCGA_seg}}
\end{figure}

For TCGA datasets, since the profiles are rather long, we did not generate annotations using the consensus approach. We apply some of the methods on this dataset and compared their segmentation results (Fig. \ref{fig:TCGA_seg}). Again, we can see that iSeg identified most of the significant peaks. 
DNACopy performs well overall, but tends to miss single-point peaks while other methods did not perform as expected.

We compare the computational time of iSeg with those of the other methods. Table \ref{tab1} shows that iSeg is the fastest method for all three data sets. It is worth noting that for very long profiles (length 100000), iSeg takes much less time than the other methods. The use of dynamic programming and a power factor makes the initial scanning of a profile fairly fast. 
The long profiles contain similar amount of data points that are signals (as opposed to background or noise) as the shorter profiles. The time spent on dealing with potentially significant segments is roughly the same between the two types of profiles. As a result, the overall running time of iSeg for the long profiles does not increase so much as the other methods. In summary, iSeg runs faster than the other methods, and much faster for profiles with sparse signals.

\section*{\huge Discussion}
In this study, we designed an efficient method, iSeg, for the segmentation of large-scale genomic profiles. When compared with existing methods using both simulated and experimental data, iSeg shows superior accuracy and speed. iSeg performs equally well when tested on very long profiles, making it suitable for real-time, online applications involving large-scale genomic datasets. In this study, we have assumed that the data follow a normal distribution. The algorithm is not limited to this distribution and other hypothesis tests can be used to compute p-values for the segments. It has been shown that data generated by next-generation sequencing (NGS), which has gained much popularity in genomics research, may follow a Poisson or a negative binomial
distribution \cite{auer2010statistical,marioni2008rna}. We aim to implement these two distributions in the near future. The use of dynamic programming and a power factor makes iSeg computationally more efficient when analyzing very long profiles, especially with sparse signals as noted in many NGS data sets. The refinement step identifies the exact boundaries of segments found by the scanning step. Merging allows iSeg to detect segments of any length. Together, these steps make iSeg an accurate and efficient method for segmentation of sequential data. \color{black} Methods such as VEGA \cite{morganella2010vega_3} isolate segments while accounting for serial dependence using a piece-wise smoothing model and minimization of a cost function. However, a piece-wise smoothing model could be problematic especially in the presence of sharp changes (peaks or very narrow segments) on a short interval. Such a method could miss short segments by calling a longer segment containing the short segment. This behavior could be seen in some results shown in \cite{morganella2010vega_3} when comparing to Ultrasome \cite{andersson2008segmental} and SMAP \cite{nilsson2009ultrasome_2}. iSeg overcomes such issues retaining a narrow segment, or a peak, if it is deemed significant.\color{black}  Although tested only on DNA copy number data, in principle, iSeg can be used to segment other types of genomic data, such as DNA methylation and histone modifications. Some adaptation or parameter tuning may be needed for different data types. 
\\
The statistic used in our method is very similar to the one used in \cite{jeng2010optimal}, which is based on similar model assumptions as some of the previous methods. However in \cite{jeng2010optimal}, the segments are identified using an exhaustive approach, which will not be efficient when the profiles to be segmented are very long. To speed up computation, the method in \cite{jeng2010optimal}  assumes that the segments have relatively short length, which is not true for some datasets. The algorithm designed in this study allows us to detect segments of any length with greater efficiency.
\\
The gold standard generated using the consensus approach does not guarantee that the true optimal segments will be identified. In addition, the $F_{1}$-scores may favor iSeg more as the test statistic used to generate the gold standard is not employed by the other methods. However, the statistic we used is based on model assumptions used by many existing methods and can be used in evaluating segments of length 1 or more. Some existing test statistics cannot be used for segments of length 1, which is the reason why they tend to miss such segments. Clearly, visual inspection of segmentation results shows that iSeg performs better than the other methods in this study. A natural extension of iSeg will be to compare multiple profiles simultaneously. This will be a subject for our future research.
\\
We have designed the method to make it flexible and versatile. This resulted in a number of parameters that users can tune. However, the default values work well for all the simulated and experimental datasets. In practice, to obtain satisfactory results, users are not expected to modify any parameters. \color{black}The speed of iSeg would allow us and fellow researchers to implement it as an online tool to deliver segmentation results in real-time.\color{black}
\\
\newpage
\section*{\huge Tables}
\begin{table}[h!]
\caption{\textbf{Comparison of computational times (in seconds) on simulated data and
Coriell data.\label{tab1}}}
{\begin{tabular*}{0.75\columnwidth}{llll}
\hline
\textbf{Method} & \textbf{Simulation } & \textbf{Simulation} & \textbf{Coriell}\\
& (SNR$\simeq$1.0, n=5000) & (SNR$\simeq$1.0, n=100K)  \\
\toprule
\textbf{iSeg (C++)} & 0.164 & 1.223 & 0.294 \\
\textbf{DNAcopy (R)} & 2.267 & 60.343 & 3.098 \\
\textbf{fastseg (R)} & 0.647 & 48.139 & 0.630 \\
\textbf{CGHSeg (R)} & 54.480 & 157.626 & 24.36 \\
\textbf{HMMSeg (Java)} & 0.543 & 160.790 & 0.552 \\
\hline
\end{tabular*}}%
\caption{The table summarizes total computational times required to process 10 simulated and 11
Coriell profiles.}
\end{table}
\bibliography{Paper-Gene}

\begin{thebibliography}{10}
\providecommand{\url}[1]{\texttt{#1}}
\providecommand{\urlprefix}{URL }
\expandafter\ifx\csname urlstyle\endcsname\relax
  \providecommand{\doi}[1]{doi:\discretionary{}{}{}#1}\else
  \providecommand{\doi}{doi:\discretionary{}{}{}\begingroup
  \urlstyle{rm}\Url}\fi
\providecommand{\bibAnnoteFile}[1]{%
  \IfFileExists{#1}{\begin{quotation}\noindent\textsc{Key:} #1\\
  \textsc{Annotation:}\ \input{#1}\end{quotation}}{}}
\providecommand{\bibAnnote}[2]{%
  \begin{quotation}\noindent\textsc{Key:} #1\\
  \textsc{Annotation:}\ #2\end{quotation}}
\providecommand{\eprint}[2][]{\url{#2}}

\bibitem{encode2012integrated}
Consortium EP, et~al. (2012) {An integrated encyclopedia of DNA elements in the
  human genome}.
\newblock Nature 489: 57–74.
\bibAnnoteFile{encode2012integrated}

\bibitem{KhatriSB122}
Khatri P, Sirota M, Butte AJ (2012) {Ten Years of Pathway Analysis: Current
  Approaches and Outstanding Challenges.}
\newblock PLoS Computational Biology 8.
\bibAnnoteFile{KhatriSB122}

\bibitem{pmid15475419}
Olshen AB, Venkatraman ES, Lucito R, Wigler M (2004) {Circular binary
  segmentation for the analysis of array-based DNA copy number data.}
\newblock {Biostatistics} 5: 557–72.
\bibAnnoteFile{pmid15475419}

\bibitem{jstor4541408}
Picard F, Robin S, Lebarbier E, Daudin JJ (2007) {A Segmentation/Clustering
  Model for the Analysis of Array CGH Data}.
\newblock Biometrics 63: 758–766.
\bibAnnoteFile{jstor4541408}

\bibitem{rancoita2009bayesian}
Rancoita P, Hutter M, Bertoni F, Kwee I (2009) {Bayesian DNA copy number
  analysis}.
\newblock BMC bioinformatics 10: 10.
\bibAnnoteFile{rancoita2009bayesian}

\bibitem{zhang2007modified}
Zhang NR, Siegmund DO (2007) A modified bayes information criterion with
  applications to the analysis of comparative genomic hybridization data.
\newblock Biometrics 63: 22–32.
\bibAnnoteFile{zhang2007modified}

\bibitem{Diskin2006}
Diskin SJ, Eck T, Greshock J, Mosse YP, Naylor T, et~al. ({2006}) {STAC: A
  method for testing the significance of DNA copy number aberrations across
  multiple array-CGH experiments}.
\newblock {GENOME RESEARCH} {16}: {1149–1158}.
\bibAnnoteFile{Diskin2006}

\bibitem{pmid17234643}
Venkatraman ES, Olshen AB (2007) {A faster circular binary segmentation
  algorithm for the analysis of array CGH data.}
\newblock {Bioinformatics} 23: 657–63.
\bibAnnoteFile{pmid17234643}

\bibitem{baldi1998machine}
Baldi P, Brunak S (1998).
\newblock {The machine learning approach}.
\bibAnnoteFile{baldi1998machine}

\bibitem{BaldiL01}
Baldi P, Long AD (2001) {A Bayesian framework for the analysis of microarray
  expression data: regularized t -test and statistical inferences of gene
  changes.}
\newblock Bioinformatics 17: 509–519.
\bibAnnoteFile{BaldiL01}

\bibitem{HMMSeg}
Day N (2007).
\newblock {HMMSeg}.
\newblock \urlprefix\url{http://noble.gs.washington.edu/proj/hmmseg}.
\bibAnnoteFile{HMMSeg}

\bibitem{PicardLBR11}
Picard F, Lebarbier E, Budinskà E, Robin S (2011) {Joint segmentation of
  multivariate Gaussian processes using mixed linear models.}
\newblock Computational Statistics \& Data Analysis 55: 1160–1170.
\bibAnnoteFile{PicardLBR11}

\bibitem{jeng2010optimal}
Jeng XJ, Cai TT, Li H (2010) {Optimal sparse segment identification with
  application in copy number variation analysis}.
\newblock Journal of the American Statistical Association 105: 1156–1166.
\bibAnnoteFile{jeng2010optimal}

\bibitem{tcai}
Cai T, Jeng J, Li H (2012) {Robust detection and identification of sparse
  segments in ultra-high dimensional data}.
\newblock Journal of the Royal Statistical Society Series B*: 773–797.
\bibAnnoteFile{tcai}

\bibitem{wang2007penncnv}
Wang K, Li M, Hadley D, Liu R, Glessner J, et~al. (2007) {PennCNV: an
  integrated hidden Markov model designed for high-resolution copy number
  variation detection in whole-genome SNP genotyping data}.
\newblock Genome research 17: 1665–1674.
\bibAnnoteFile{wang2007penncnv}

\bibitem{sen1975tests}
Sen A, MS S (1975) {On tests for detecting change in mean}.
\newblock The Annals of Statistics 3: 98–108.
\bibAnnoteFile{sen1975tests}

\bibitem{Picard2005}
Picard F, Robin S, Lavielle M, Vaisse C, Daudin J (2005) A statistical approach
  for array cgh data analysis.
\newblock BMC Bioinformatics 6: 27.
\bibAnnoteFile{Picard2005}

\bibitem{chen2009statistical}
Chen J, Wang YP (2009) A statistical change point model approach for the
  detection of dna copy number variations in array cgh data.
\newblock Computational Biology and Bioinformatics, IEEE/ACM Transactions on 6:
  529–541.
\bibAnnoteFile{chen2009statistical}

\bibitem{chen2011bayesian}
Chen J, Yiğiter A, Chang KC (2011) A bayesian approach to inference about a
  change point model with application to dna copy number experimental data.
\newblock Journal of Applied Statistics 38: 1899–1913.
\bibAnnoteFile{chen2011bayesian}

\bibitem{Niu2012}
Niu Y, Zhang H (2012) The screening and ranking algorithm to detect dna copy
  number variations.
\newblock Annals of Applied Statistics 6: 1306–1326.
\bibAnnoteFile{Niu2012}

\bibitem{Yao1993}
YAO Q ({1993}) {TESTS FOR CHANGE-POINTS WITH EPIDEMIC ALTERNATIVES}.
\newblock {BIOMETRIKA} {80}: {179–191}.
\bibAnnoteFile{Yao1993}

\bibitem{killick2011changepoint}
Killick R, Eckley IA (2011) Changepoint: an r package for changepoint analysis.
\newblock R package version 06, URL http://CRAN R-project org/package=
  changepoint .
\bibAnnoteFile{killick2011changepoint}

\bibitem{cleynen2013segmentor3isback}
Cleynen A, Koskas M, Lebarbier E, Rigaill G, Robin S (2013) Segmentor3isback:
  an r package for the fast and exact segmentation of seq-data.
\newblock In: The R User Conference, University of Castilla-La Mancha,
  Albacete, Spain.
\bibAnnoteFile{cleynen2013segmentor3isback}

\bibitem{Marioni2006}
Marioni JC, Thorne NP, Tavar� S (2006) Biohmm: a heterogeneous hidden markov
  model for segmenting array cgh data.
\newblock Bioinformatics 22: 1144–1146.
\bibAnnoteFile{Marioni2006}

\bibitem{Stjernqvist2007}
Stjernqvist S, Ryd�n T, Sk�ld M, Staaf J (2007) Continuous-index hidden
  markov modelling of array cgh copy number data.
\newblock Bioinformatics 23: 1006–1014.
\bibAnnoteFile{Stjernqvist2007}

\bibitem{jaschek2009spatial}
Jaschek R, Tanay A (2009) {Spatial clustering of multivariate genomic and
  epigenomic information}.
\newblock In: {Research in Computational Molecular Biology}. Springer, p.
  170–183.
\bibAnnoteFile{jaschek2009spatial}

\bibitem{hoffman2012unsupervised}
Hoffman MM, Buske OJ, Wang J, Weng Z, Bilmes JA, et~al. (2012) {Unsupervised
  pattern discovery in human chromatin structure through genomic segmentation}.
\newblock Nature methods 9: 473–476.
\bibAnnoteFile{hoffman2012unsupervised}

\bibitem{hoffman2012integrative}
Hoffman MM, Ernst J, Wilder SP, Kundaje A, Harris RS, et~al. (2012)
  {Integrative annotation of chromatin elements from ENCODE data}.
\newblock Nucleic acids research : gks1284.
\bibAnnoteFile{hoffman2012integrative}

\bibitem{hupe2004analysis}
Hupé P, Stransky N, Thiery JP, Radvanyi F, Barillot E (2004) Analysis of array
  cgh data: from signal ratio to gain and loss of dna regions.
\newblock Bioinformatics 20: 3413–3422.
\bibAnnoteFile{hupe2004analysis}

\bibitem{ben2008fast}
Ben-Yaacov E, Eldar YC (2008) A fast and flexible method for the segmentation
  of acgh data.
\newblock Bioinformatics 24: i139–i145.
\bibAnnoteFile{ben2008fast}

\bibitem{Tibshirani2008}
Tibshirani R, Wang P (2008) Spatial smoothing and hot spot detection for cgh
  data using the fused lasso.
\newblock Biostatistics 9: 18–29.
\bibAnnoteFile{Tibshirani2008}

\bibitem{hu2007exploiting}
Hu J, Gao JB, Cao Y, Bottinger E, Zhang W (2007) Exploiting noise in array cgh
  data to improve detection of dna copy number change.
\newblock Nucleic acids research 35: e35.
\bibAnnoteFile{hu2007exploiting}

\bibitem{nilsson2009ultrasome}
Nilsson B, Johansson M, Al-Shahrour F, Carpenter AE, Ebert BL (2009) Ultrasome:
  efficient aberration caller for copy number studies of ultra-high resolution.
\newblock Bioinformatics 25: 1078–1079.
\bibAnnoteFile{nilsson2009ultrasome}

\bibitem{Morganella2010}
Morganella S, Cerulo L, Viglietto G, Ceccarelli M (2010) Vega: variational
  segmentation for copy number detection.
\newblock Bioinformatics 26: 3020–3027.
\bibAnnoteFile{Morganella2010}

\bibitem{kharchenko2008design}
Kharchenko P, Tolstorukov M, Park P (2008) {Design and analysis of ChIP-seq
  experiments for DNA-binding proteins}.
\newblock Nature biotechnology 26: 1351–1359.
\bibAnnoteFile{kharchenko2008design}

\bibitem{park2008experimental}
Park P (2008) {Experimental design and data analysis for array comparative
  genomic hybridization}.
\newblock Cancer investigation 26: 923–928.
\bibAnnoteFile{park2008experimental}

\bibitem{lai2005comparative}
Lai WR, Johnson MD, Kucherlapati R, Park PJ (2005) Comparative analysis of
  algorithms for identifying amplifications and deletions in array cgh data.
\newblock Bioinformatics 21: 3763–3770.
\bibAnnoteFile{lai2005comparative}

\bibitem{willenbrock2005comparison}
Willenbrock H, Fridlyand J (2005) {A comparison study: applying segmentation to
  array CGH data for downstream analyses}.
\newblock Bioinformatics 21: 4084–4091.
\bibAnnoteFile{willenbrock2005comparison}

\bibitem{zhang2012reconstructing}
Zhang Z, Lange K, Sabatti C (2012) Reconstructing dna copy number by joint
  segmentation of multiple sequences.
\newblock BMC bioinformatics 13: 205.
\bibAnnoteFile{zhang2012reconstructing}

\bibitem{zhou2013multisample}
Zhou X, Yang C, Wan X, Zhao H, Yu W (2013) Multisample acgh data analysis via
  total variation and spectral regularization.
\newblock Computational Biology and Bioinformatics, IEEE/ACM Transactions on
  10: 230–235.
\bibAnnoteFile{zhou2013multisample}

\bibitem{Zhang2010}
Zhang Q, Ding L, Larson DE, Koboldt DC, McLellan MD, et~al. (2010) Cmds: a
  population-based method for identifying recurrent dna copy number aberrations
  in cancer from high-resolution data.
\newblock Bioinformatics 26: 464–469.
\bibAnnoteFile{Zhang2010}

\bibitem{baladandayuthapani2010bayesian}
Baladandayuthapani V, Ji Y, Talluri R, Nieto-Barajas LE, Morris JS (2010)
  Bayesian random segmentation models to identify shared copy number
  aberrations for array cgh data.
\newblock Journal of the American Statistical Association 105.
\bibAnnoteFile{baladandayuthapani2010bayesian}

\bibitem{Pique-Regi2009}
Pique-Regi R, Ortega A, Asgharzadeh S (2009) Joint estimation of copy number
  variation and reference intensities on multiple dna arrays using gada.
\newblock Bioinformatics 25: 1223–1230.
\bibAnnoteFile{Pique-Regi2009}

\bibitem{Wiel2009}
van~de Wiel MA, Brosens R, Eilers PHC, Kumps C, Meijer GA, et~al. (2009)
  Smoothing waves in array cgh tumor profiles.
\newblock Bioinformatics 25: 1099–1104.
\bibAnnoteFile{Wiel2009}

\bibitem{Picard2011a}
Picard F, Lebarbier E, Hoebeke M, Rigaill G, Thiam B, et~al. (2011) Joint
  segmentation, calling, and normalization of multiple cgh profiles.
\newblock Biostatistics 12: 413–428.
\bibAnnoteFile{Picard2011a}

\bibitem{Guttman2007}
Guttman M, Mies C, Dudycz-Sulicz K, Diskin SJ, Baldwin DA, et~al. (2007)
  Assessing the significance of conserved genomic aberrations using high
  resolution genomic microarrays.
\newblock PLoS Genet 3: e143.
\bibAnnoteFile{Guttman2007}

\bibitem{beroukhim2007assessing}
Beroukhim R, Getz G, Nghiemphu L, Barretina J, Hsueh T, et~al. (2007) Assessing
  the significance of chromosomal aberrations in cancer: methodology and
  application to glioma.
\newblock Proceedings of the National Academy of Sciences 104: 20007–20012.
\bibAnnoteFile{beroukhim2007assessing}

\bibitem{Shah2007}
Shah SP, Lam WL, Ng RT, Murphy KP (2007) Modeling recurrent dna copy number
  alterations in array cgh data.
\newblock Bioinformatics 23: i450–i458.
\bibAnnoteFile{Shah2007}

\bibitem{Zhang2010b}
Zhang NR, Senbabaoglu Y, Li JZ (2010) Joint estimation of dna copy number from
  multiple platforms.
\newblock Bioinformatics 26: 153–160.
\bibAnnoteFile{Zhang2010b}

\bibitem{Zhang2010a}
Zhang NR, Siegmund DO, Ji H, Li JZ (2010) Detecting simultaneous changepoints
  in multiple sequences.
\newblock Biometrika 97: 631–645.
\bibAnnoteFile{Zhang2010a}

\bibitem{Nowak2011}
Nowak G, Hastie T, Pollack JR, Tibshirani R (2011) A fused lasso latent feature
  model for analyzing multi-sample acgh data.
\newblock Biostatistics 12: 776–791.
\bibAnnoteFile{Nowak2011}

\bibitem{roy2013evaluation}
Roy S, Reif AM (2013) Evaluation of calling algorithms for array-cgh.
\newblock Frontiers in genetics 4.
\bibAnnoteFile{roy2013evaluation}

\bibitem{brodsky1993nonparametric}
Brodsky BE, Darkhovsky BS (1993) Nonparametric methods in change point
  problems.
\newblock 243. Springer.
\bibAnnoteFile{brodsky1993nonparametric}

\bibitem{benjamini1995controlling}
Benjamini Y, Hochberg Y (1995) {Controlling the false discovery rate: a
  practical and powerful approach to multiple testing}.
\newblock Journal of the Royal Statistical Society Series B (Methodological) :
  289–300.
\bibAnnoteFile{benjamini1995controlling}

\bibitem{pmid17384021}
Day N, Hemmaplardh A, Thurman RE, Stamatoyannopoulos JA, Noble WS (2007)
  {Unsupervised segmentation of continuous genomic data.}
\newblock {Bioinformatics} 23: 1424–6.
\bibAnnoteFile{pmid17384021}

\bibitem{Chinchor92}
Chinchor N (1992) {MUC-4 Evaluation Metrics}.
\newblock In: {Proceedings of the Fourth Message Understanding Conference}. p.
  22–29.
\bibAnnoteFile{Chinchor92}

\bibitem{sokolova2006beyond}
Sokolova M, Japkowicz N, Szpakowicz S (2006) Beyond accuracy, f-score and roc:
  a family of discriminant measures for performance evaluation.
\newblock In: AI 2006: Advances in Artificial Intelligence, Springer. p.
  1015–1021.
\bibAnnoteFile{sokolova2006beyond}

\bibitem{snijders2001assembly}
Snijders A, Nowak N, Segraves R, Blackwood S, Brown N, et~al. (2001) {Assembly
  of microarrays for genome-wide measurement of DNA copy number by CGH}.
\newblock Nature genetics 29: 263–264.
\bibAnnoteFile{snijders2001assembly}

\bibitem{wang2005method}
Wang P, Kim Y, Pollack J, Narasimhan B, Tibshirani R (2005) A method for
  calling gains and losses in array cgh data.
\newblock Biostatistics 6: 45–58.
\bibAnnoteFile{wang2005method}

\bibitem{auer2010statistical}
Auer PL, Doerge R (2010) {Statistical design and analysis of RNA sequencing
  data}.
\newblock Genetics 185: 405–416.
\bibAnnoteFile{auer2010statistical}

\bibitem{marioni2008rna}
Marioni JC, Mason CE, Mane SM, Stephens M, Gilad Y (2008) {RNA-seq: an
  assessment of technical reproducibility and comparison with gene expression
  arrays}.
\newblock Genome research 18: 1509–1517.
\bibAnnoteFile{marioni2008rna}

\bibitem{morganella2010vega_3}
Morganella S, Cerulo L, Viglietto G, Ceccarelli M (2010) Vega: Variational
  segmentation for copy number detection.
\newblock Bioinformatics 26: 3020–3027.
\bibAnnoteFile{morganella2010vega_3}

\bibitem{andersson2008segmental}
Andersson R, Bruder CE, Piotrowski A, Menzel U, Nord H, et~al. (2008) A
  segmental maximum a posteriori approach to genome-wide copy number profiling.
\newblock Bioinformatics 24: 751–758.
\bibAnnoteFile{andersson2008segmental}

\bibitem{nilsson2009ultrasome_2}
Nilsson B, Johansson M, Al-Shahrour F, Carpenter AE, Ebert BL (2009) Ultrasome:
  efficient aberration caller for copy number studies of ultra-high resolution.
\newblock Bioinformatics 25: 1078–1079.
\bibAnnoteFile{nilsson2009ultrasome_2}

\end{thebibliography}
\end{document}